# Generalizing second order macroscopic models via relative velocity of disturbances propagation in traffic flow

Y.A. Kholodov[a,*], A.E. Alekseenko[a,b], A.S. Kholodov[b,c], A.N. Karachev[a,b], A.A. Kurzhanskiy[d]

[a] *Innopolis University, Innopolis, Tatarstan, 420500, Russia*
[b] *Institute for Computer Aided Design of RAS, Moscow, 123056, Russia*
[c] *Moscow Institute of Physics and Technology, Dolgoprudny, Moscow Region, 141707, Russia*
[d] *University of California, Berkeley, CA 94720-1770, United States*
[*] *Corresponding author: ya.kholodov@innopolis.ru*

**Abstract**

In this paper, we present a new approach that generalizes the existing second-order hydrodynamic traffic models. In the proposed approach, we use the expression for the relative velocity of disturbances propagation in traffic flow. We show that properties of any phenomenological model are fully defined by how the relative velocity of disturbances propagation expressed in the model. We verify the proposed approach through simulations of the Interstate 580 freeway segment in California, USA, with traffic measurements from the Performance Measurement System (PeMS).

The work was supported by grants RSCF 14-11-00877.

## 1. Introduction

An extensive development of gas dynamics (generalized conservation laws, stable differencing schemes) has taken place in the 1950s. At the same time, first macroscopic (hydrodynamical) models emerged, in which traffic flow was likened to flow of "motivated" compressible liquid. In the Lighthill–Whitham–Richards (LWR) model (Lighthill and Whitham,1955; Richards,1956; Whitham,1974) the traffic flow is described by the law of conservation of vehicles. This model postulates the unique dependence of traffic flow on traffic density, which is called a fundamental diagram. In recent years, the class of macroscopic models has expanded significantly. Nowadays, the traffic flow is described using system of non-linear hyperbolic second-order PDEs (for traffic density and speed), in various formulations (Payne,1971; Daganzo,1995; Papageorgiou,1998; Aw and Rascle,2000; Zhang,2002,2003; Siebel and Mauser,2006a,b), which do not assume unique dependence of speed on density.

In this paper, we present a new approach that generalizes the existing second-order traffic models. In the existing second-order models, the traffic flow is described using system of non-linear hyperbolic second-order PDEs (for traffic density and speed) (Payne,1971; Daganzo,1995; Papageorgiou,1998; Aw and Rascle,2000; Zhang,2002,2003; Siebel and Mauser,2006a,b), which account for the dependence of traffic flow (or speed) on its density in different ways. We introduce the relative velocity of disturbances propagation in the traffic flow that unifies all second-order hydrodynamical models.

The relative velocity of disturbances propagation in the traffic flow for each roadway segment can be obtained empirically using observed data from traffic detectors. Furthermore, knowing the value of the relative velocity, we can do without a fundamental diagram. We verify the proposed approach through numerical experiments using typical traffic detector data from the Performance Measurement System (PeMS) (California Department of Transportation,2012), on the I-580 freeway in California, USA. Simulations show different second-order models using the same state equation produce the same results.

The rest of the paper is organized as follows. Section 2 has three parts: Subsection 2.1 describes the generalized second-order model and shows that the existing second-order models are special cases of the proposed one; Subsection 2.2 presents the numerical approximation for model simulation. Section 3 discusses simulation results, which support the proposition that different second-order models with the same expression for the relative velocity of disturbances propagation behave the same way. Finally, Section 4 concludes the paper.

## 2. Generalized model

### 2.1 Model equation system

The Lighthill-Whithem-Richards (LWR) model (Lighthill and Whitham,1955; Richards,1956) assumes a unique relationship between traffic speed $v(t,x)$ and density $\rho(t, x)$: $v(t,x) = V(\rho(t,x))$, and the vehicle conservation law. Here, $\rho(t,x)$ denotes the number of vehicles per unit length of the roadway at moment $t$, around the point with coordinate $x$; and $v(t, x)$ denotes mean traffic speed at time $t$ around the point with coordinate $x$. The function $V(\rho)$ is non-increasing: $\partial V/\partial \rho \leq 0$. We denote the flow (i.e., the number of vehicles passing a reference point in unit time) as $Q(\rho) = \rho V(\rho)$. The function $Q(\rho)$ is usually referred to as the fundamental diagram (although sometimes this name is reserved for $V(\rho)$).

The vehicle conservation law in the LWR model is expressed in the differential form of continuity equation with zero right-hand side:

$$\frac{\partial \rho}{\partial t} + \frac{\partial(\rho v)}{\partial x} = \frac{\partial \rho}{\partial t} + \frac{\partial(\rho V(\rho))}{\partial x} = \frac{\partial \rho}{\partial t} + \frac{\partial Q(\rho)}{\partial x} = 0 \qquad (1)$$

If we account for variations in vehicle number, for example, due to lane changes or freeway entrances and exits, the Eq. (1) takes the form:

$$\frac{\partial \rho}{\partial t} + \frac{\partial Q(\rho)}{\partial x} = f_0, \qquad (2)$$

where $f_0$ is the number of vehicles entering (positive) or leaving (negative) per unit time. The Eq. (2) by itself is not sufficient for the adequate description of all the phases of traffic (Daganzo,1995). To correct this, we use the differential transformation of the conservation law (Godunov and Romenskii,2003). We multiply (2) by $\partial v/\partial \rho$:

$$\frac{\partial v}{\partial \rho}\left(\frac{\partial \rho}{\partial t} + \frac{\partial Q}{\partial x}\right) = \frac{\partial v}{\partial \rho}\frac{\partial \rho}{\partial t} + \frac{\partial v}{\partial \rho}\left(\rho\frac{\partial v}{\partial x} + v\frac{\partial \rho}{\partial x}\right) = \frac{\partial v}{\partial \rho} f_0,$$

and we get

$$\frac{\partial v}{\partial t} + v\frac{\partial v}{\partial x} + \rho\frac{\partial v}{\partial \rho}\frac{\partial v}{\partial x} = \frac{\partial v}{\partial t} + \left(v + \rho\frac{\partial v}{\partial \rho}\right)\frac{\partial v}{\partial x} = \frac{\partial v}{\partial \rho} f_0 \qquad (3)$$

If we take the derivative of $\partial Q/\partial \rho$ keeping in mind that $Q(v,\rho)$ is a function of two variables:

$$\frac{\partial Q}{\partial \rho} = \frac{\partial(v\rho)}{\partial \rho} = v + \rho\frac{\partial v}{\partial \rho},$$

we can rewrite equations (2),(3) in the next simple form

$$\begin{cases} \dfrac{\partial \rho}{\partial t} + \dfrac{\partial Q}{\partial x} = \dfrac{\partial \rho}{\partial t} + \left(v + \rho\dfrac{\partial v}{\partial \rho}\right)\dfrac{\partial \rho}{\partial x} = \dfrac{\partial \rho}{\partial t} + \left(\dfrac{\partial Q}{\partial \rho}\right)\dfrac{\partial \rho}{\partial x} = f_0, \\ \dfrac{\partial v}{\partial t} + \left(v + \rho\dfrac{\partial v}{\partial \rho}\right)\dfrac{\partial v}{\partial x} = \dfrac{\partial v}{\partial t} + \left(\dfrac{\partial Q}{\partial \rho}\right)\dfrac{\partial v}{\partial x} = \dfrac{\partial v}{\partial \rho} f_0. \end{cases} \qquad (4)$$

Thus, we constructed a second-order macroscopic model, which does not assume unique dependency between traffic speed and density, as was in the LWR. However, in (4) we need to evaluate the dependence of $\partial v/\partial \rho$ in velocity equation to integrate it. At the same time, we can rid it off in the density equation. Later in Subsection 2.3, we discuss how to do it straightforwardly.

The right-hand side $f_0$ of the first equation in (4) accounts for the number of vehicles incoming (positive) or leaving (negative) per unit time. Looking at the right-hand side of the velocity equation in (4), we see the impact of these vehicles on traffic speed.

The number of left and right boundary conditions depends on the signs of the eigenvalues of the system (4): $\lambda_1 = v$, $\lambda_2 = v + \rho \frac{\partial v}{\partial \rho}$. Because $\lambda_1 = v$ always $\geq 0$ we need to know the sign of $\lambda_2 = v + \rho \frac{\partial v}{\partial \rho}$. At a freeway entrance, their number is either two, if both $\lambda_{1,2} > 0$, either one, if $\lambda_1 > 0$, $\lambda_2 \leq 0$, or zero, if $\lambda_1 = 0$, $\lambda_2 \leq 0$. The opposite is true for a freeway exit: zero boundary conditions, if both $\lambda_{1,2} \geq 0$, and one, if $\lambda_1 \geq 0$, $\lambda_2 < 0$. Therefore, we can use time-dependent traffic flow $Q(t)$ and speed $v(t)$ as boundary conditions. Besides boundary conditions, we set the following initial conditions:

$$\rho(x,0) = \rho_0(x), v(x,0) = v_0(x).$$

Existing macroscopic second-order models describe traffic in various formulations (Payne,1971; Daganzo,1995; Papageorgiou,1998; Aw and Rascle,2000; Zhang,2002,2003; Siebel and Mauser,2006a,b). These models differ in the way they describe the dependency of flow (or speed) on density. Now, we will discuss the most popular second-order models and show how they turn out to the special cases of proposed model (4).

We start with the Payne–Whitham model (Payne,1971; Whitham,1974) that has zero right-hand side:

$$\begin{cases} \frac{\partial \rho}{\partial t} + \frac{\partial Q}{\partial x} = 0 \\ \frac{\partial v}{\partial t} + v \frac{\partial v}{\partial x} + \frac{c_0^2}{\rho} \frac{\partial \rho}{\partial x} = 0 \end{cases} \quad (5)$$

and generalize it for the arbitrary form of the dependency between the pressure and the density:

$$P(\rho) = \int_0^\rho c(\tilde{\rho})^2 \, d\tilde{\rho} \iff \partial P / \partial \rho = c(\rho)^2 \quad (6)$$

With $P(\rho)$ defined in (6), we can rewrite model (5) as (Zhang,2002):

$$\begin{cases} \frac{\partial \rho}{\partial t} + \frac{\partial Q}{\partial x} = \frac{\partial \rho}{\partial t} + \left(\frac{\partial Q}{\partial \rho}\right) \frac{\partial \rho}{\partial x} = \frac{\partial \rho}{\partial t} + v \frac{\partial \rho}{\partial x} + \rho \frac{\partial v}{\partial x} = 0, \\ \frac{\partial v}{\partial t} + v \frac{\partial v}{\partial x} + \frac{c(\rho)^2}{\rho} \frac{\partial \rho}{\partial x} = \frac{\partial v}{\partial t} + v \frac{\partial v}{\partial x} + \frac{1}{\rho} \frac{\partial P}{\partial \rho} \frac{\partial \rho}{\partial x} = \frac{\partial v}{\partial t} + v \frac{\partial v}{\partial x} + \frac{1}{\rho} \frac{\partial P}{\partial x} = 0, \end{cases} \quad (7)$$

which generalizes (5).

System (7) could also be rewritten in the divergent form:

$$\begin{cases} \dfrac{\partial \rho}{\partial t} + \dfrac{\partial (\rho v)}{\partial x} = 0 \\ \dfrac{\partial (\rho v)}{\partial t} + \dfrac{\partial \left(\rho v^2 + P(\rho)\right)}{\partial x} = 0 \end{cases} \tag{8}$$

Now we need to find a way to express pressure $P(\rho) = \int_0^\rho c(\tilde{\rho})^2 \, d\tilde{\rho}$ or the pressure derivative by density $\partial P/\partial \rho = c(\rho)^2$ through calculated variables, speed $v$, and density ρ. First, we express partial derivative ∂P/∂ρ:

$$\frac{\partial P}{\partial \rho} = \frac{\partial P}{\partial x}\frac{\partial x}{\partial \rho} = \frac{P_x}{\rho_x} = \left(\rho_x = -\frac{\rho_t}{Q_\rho}\right) = -\frac{Q_\rho}{\rho_t} P_x \tag{9}$$

Then, we express ∂P/∂x:

$$-P_x = -\frac{\partial P}{\partial x} = \frac{\partial Q}{\partial t} + \frac{\partial (Qv)}{\partial x} = \frac{\partial Q}{\partial t} + \frac{\partial (Qv)}{\partial Q}\frac{\partial Q}{\partial x} = \left(\frac{\partial Q}{\partial x} = -\frac{\partial \rho}{\partial t}\right) = \frac{\partial Q}{\partial t} - \frac{\partial (Qv)}{\partial Q}\frac{\partial \rho}{\partial t}$$

and substitute it into Eq. (9):

$$\frac{\partial P}{\partial \rho} = -\frac{Q_\rho}{\rho_t} P_x = \frac{Q_\rho}{\rho_t}\left(\frac{\partial Q}{\partial t} - \frac{\partial (Qv)}{\partial Q}\frac{\partial \rho}{\partial t}\right) = \frac{\partial Q}{\partial \rho}\left(\frac{\partial Q}{\partial \rho} - \frac{\partial (Qv)}{\partial Q}\right) = \left(\frac{\partial Q}{\partial \rho}\right)^2 - \frac{\partial (Qv)}{\partial \rho}$$

Taking $\dfrac{\partial Q}{\partial \rho}$ and $\dfrac{\partial (Qv)}{\partial \rho}$ from the known relations:

$$\left(\frac{\partial Q}{\partial \rho}\right)^2 = \left(\frac{\partial (\rho v)}{\partial \rho}\right)^2 = \left(v + \rho\frac{\partial v}{\partial \rho}\right)^2 = v^2 + 2\rho v\frac{\partial v}{\partial \rho} + \rho^2\left(\frac{\partial v}{\partial \rho}\right)^2$$

$$\frac{\partial (Qv)}{\partial \rho} = \frac{\partial (\rho v^2)}{\partial \rho} = v^2 + \rho\frac{\partial (v^2)}{\partial \rho} = v^2 + 2\rho v\frac{\partial v}{\partial \rho}$$

we obtain the final equation for ∂P/∂ρ = $c^2(\rho)$:

$$\frac{\partial P}{\partial \rho} = c(\rho)^2 = \left(\frac{\partial Q}{\partial \rho}\right)^2 - \frac{\partial (Qv)}{\partial \rho} = \rho^2\left(\frac{\partial v}{\partial \rho}\right)^2 \tag{10}$$

Using (10) in the second equation of (7):

$$\begin{cases} \dfrac{\partial \rho}{\partial t} + \dfrac{\partial Q}{\partial x} = \dfrac{\partial \rho}{\partial t} + \left(\dfrac{\partial Q}{\partial \rho}\right)\dfrac{\partial \rho}{\partial x} = \dfrac{\partial \rho}{\partial t} + v\dfrac{\partial \rho}{\partial x} + \rho\dfrac{\partial v}{\partial x} = 0 \\ \dfrac{\partial v}{\partial t} + v\dfrac{\partial v}{\partial x} + \dfrac{c^2}{\rho}\dfrac{\partial \rho}{\partial x} = \dfrac{\partial v}{\partial t} + v\dfrac{\partial v}{\partial x} + \rho\dfrac{\partial v}{\partial \rho}\dfrac{\partial v}{\partial \rho}\dfrac{\partial \rho}{\partial x} = \dfrac{\partial v}{\partial t} + \left(v + \rho\dfrac{\partial v}{\partial \rho}\right)\dfrac{\partial v}{\partial x} = 0 \end{cases} \quad (11)$$

Substituting $\dfrac{\partial Q}{\partial \rho} = v + \rho\dfrac{\partial v}{\partial \rho}$ into (11), we arrive at:

$$\begin{cases} \dfrac{\partial \rho}{\partial t} + \dfrac{\partial Q}{\partial x} = \dfrac{\partial \rho}{\partial t} + \left(\dfrac{\partial Q}{\partial \rho}\right)\dfrac{\partial \rho}{\partial x} = 0 \\ \dfrac{\partial v}{\partial t} + \left(v + \rho\dfrac{\partial v}{\partial \rho}\right)\dfrac{\partial v}{\partial x} = \dfrac{\partial v}{\partial t} + \left(\dfrac{\partial Q}{\partial \rho}\right)\dfrac{\partial v}{\partial x} = 0 \end{cases} \quad (12)$$

which is the proposed model (4) with zero right-hand side.

**Remark.** It should be noted that (Zhang,2002) was the first to propose the expression for the relative velocity of disturbances propagation in the form: $c(\rho) = \rho\dfrac{\partial V}{\partial \rho}$. This differs from (11) in that instead of the observed speed $v$, it used equilibrium velocity $V(\rho)$. The expression for the relative velocity of disturbances propagation $c(\rho) = \rho\dfrac{\partial V}{\partial \rho}$ in (Zhang,2002) was derived from the car-following model (Helbing,2001), while we obtained our expression $c(\rho) = \rho\dfrac{\partial v}{\partial \rho}$ from system (8), with no additional restrictions on the form of $v(\rho)$.

Several "defects" of the model (5) proposed (Payne,1971) were discussed in the paper (Daganzo,1995). Particular, it was mention that with strong spatial disturbances in initial conditions, negative values of velocities or density might appear in the solution. Also, according to the model the traffic is influenced by vehicles going behind it that is not possible in the case of a one-lane road. These inconsistencies appear because initially the equation system (5) or (7) was used to simulate barotropic gas that is isotropic, and all directions of motion for barotropic gas are equally probable.

Mathematically this relates to the fact that the relative velocity of perturbation propagation $c(\rho)$ is squared in the second equations (7) and (8) thus these equations are not sensitive to the relative velocity sign and do not care of perturbation directions.

To remove this inconsistency in the equation system (11), we turned from $c(\rho)^2$ to $c(\rho) = \rho\dfrac{\partial v}{\partial \rho}$ and obtained the proposed second-order macroscopic model (4) with zero right-hand side. It should also be noted once we found the expression for the relative velocity of disturbances propagation in the traffic flow $c(\rho) = \rho\dfrac{\partial v}{\partial \rho}$ the eigenvalues of the isotropic system of equations (7) $\lambda_{1,2} = v \pm c(\rho)$ are transformed to the form $\lambda_1 = v$

, $\lambda_2 = v + c(\rho) \leq v$ in the system of equations (11) and no longer exceed traffic velocity.

We state out the obtained results in the form of the following theorem.

**Theorem 1.** *Any macroscopic second-order hyperbolic model of Aw-Rascle or Payne-Witham class can be formulated in the following form:*

$$\begin{cases} \dfrac{\partial \rho}{\partial t} + \dfrac{\partial Q}{\partial x} = f_0, \\ \dfrac{\partial v}{\partial t} + \left( \dfrac{\partial Q}{\partial \rho} \right) \dfrac{\partial v}{\partial x} = \dfrac{\partial v}{\partial \rho} f_0 + f_1, \end{cases} \quad (13)$$

*by means of the relative velocity of disturbances propagation in the form* $c(\rho) = \rho \dfrac{\partial v}{\partial \rho}$ *in the expression for* $\dfrac{\partial Q}{\partial \rho} = v + \rho \dfrac{\partial v}{\partial \rho} = v + c(\rho)$. *The term $f_0$ in the right-hand side of (13) accounts for the number of vehicles incoming or leaving in unit time. The term $f_1$ plays the role of a relaxation term if necessary.*

Now, let's look how this approach works in other well-known second order macroscopic models. The work (Zhang, 2002) proposes model defined by the following system:

$$\begin{cases} \dfrac{\partial \rho}{\partial t} + \dfrac{\partial Q}{\partial x} = 0 \\ \dfrac{\partial v}{\partial t} + v \dfrac{\partial v}{\partial x} + c(\rho) \dfrac{\partial v}{\partial x} = 0 \end{cases} \quad (14)$$

here the relative velocity of disturbances propagation is expressed through the derivative of the equilibrium speed: $c(\rho) = \rho \dfrac{\partial V}{\partial \rho}$.

Note that the equation system (14) is almost identical to the system proposed in (Aw and Rascle, 2000) if we take second equation in non-divergent form. The difference is in notation: in (Aw and Rascle, 2000) smooth function $p(\rho)$ is used instead of the equilibrium speed $V(\rho)$:

$$\begin{cases} \dfrac{\partial \rho}{\partial t} + \dfrac{\partial Q}{\partial x} = 0 \\ \dfrac{\partial v}{\partial t} + \left( v - \rho \dfrac{\partial p(\rho)}{\partial \rho} \right) \dfrac{\partial v}{\partial x} = \dfrac{\partial v}{\partial t} + \left( v + \rho \dfrac{\partial V(\rho)}{\partial \rho} \right) \dfrac{\partial v}{\partial x} = 0 \end{cases} \quad (15)$$

From the physical standpoint, the use of the equilibrium speed $V(\rho)$ in (15) instead of an arbitrary smooth increasing function $p(\rho) \sim \rho^\gamma$, $\gamma > 0$ is more justified (Zhang, 2002): function $V(\rho)$ cannot be smooth in points of transition between different traffic phases.

The system similar to (14), but with additional relaxation term $\beta(\rho, v)(V(\rho) - v)$ in the right-hand side, was used in (Siebel and Mauser, 2006a,b):

$$\begin{cases} \dfrac{\partial \rho}{\partial t} + \dfrac{\partial Q}{\partial x} = 0 \\ \dfrac{\partial v}{\partial t} + \left(v + \rho \dfrac{\partial V(\rho)}{\partial \rho}\right) \dfrac{\partial v}{\partial x} = \beta(\rho, v)(V(\rho) - v) \end{cases} \quad (16)$$

where the equilibrium speed is expressed as $V(\rho) = V_{max}\left(1 - \left(\dfrac{\rho}{\rho_{max}}\right)^{n_1}\right)^{n_2}$ in (Siebel and Mauser,2006a); and as $V(\rho) = V_{max}\left(1 - \exp\left(-\dfrac{\lambda}{V_{max}}\left(\dfrac{1}{\rho} - \dfrac{1}{\rho_{max}}\right)\right)\right)$ in (Siebel and Mauser,2006b).

Now we will show how model equation system in the divergent form, proposed in (Aw and Rascle,2000):

$$\begin{cases} \dfrac{\partial \rho}{\partial t} + \dfrac{\partial Q}{\partial x} = 0, \\ \dfrac{\partial (\rho(v + p(\rho)))}{\partial t} + \dfrac{\partial (\rho v(v + p(\rho)))}{\partial x} = 0, \end{cases} \quad (17)$$

can be derived from the system in form (8) using the expression for the relative velocity of disturbances propagation in the traffic flow in the form: $c(\rho) = -\rho \dfrac{\partial p(\rho)}{\partial \rho}$. First equations in both systems are identical. Thus we will focus on the second or momentum equations:

$$\dfrac{\partial(\rho v)}{\partial t} + \dfrac{\partial(\rho v^2 + P(\rho))}{\partial x} = \dfrac{\partial(\rho v)}{\partial t} + \dfrac{\partial(\rho v^2)}{\partial x} + \dfrac{\partial P(\rho)}{\partial \rho} \dfrac{\partial \rho}{\partial x} = 0 \iff$$

$$\dfrac{\partial(\rho v)}{\partial t} + \dfrac{\partial(\rho v^2)}{\partial x} + c^2(\rho) \dfrac{\partial \rho}{\partial x} = 0 \iff$$

$$\dfrac{\partial(\rho v)}{\partial t} + \dfrac{\partial(\rho v^2)}{\partial x} + \rho^2 \left(\dfrac{\partial p(\rho)}{\partial \rho}\right)^2 \dfrac{\partial \rho}{\partial x} = 0 \iff \quad (18)$$

$$\dfrac{\partial(\rho v)}{\partial t} + \dfrac{\partial(\rho v^2)}{\partial x} + \rho^2 \dfrac{\partial p(\rho)}{\partial \rho} \dfrac{\partial p(\rho)}{\partial x} = 0$$

From the first equation of (8), using $\dfrac{\partial Q}{\partial \rho} = v + c(\rho) = v - \rho \dfrac{\partial p(\rho)}{\partial \rho}$, we obtain:

$$\frac{\partial \rho}{\partial t}+\frac{\partial Q}{\partial x}=\frac{\partial \rho}{\partial t}+\frac{\partial Q}{\partial \rho}\frac{\partial \rho}{\partial x}=\frac{\partial \rho}{\partial t}+\left(v-\rho\frac{\partial p(\rho)}{\partial \rho}\right)\frac{\partial \rho}{\partial x}=0 \Leftrightarrow$$

$$\frac{\partial p(\rho)}{\partial \rho}\left(\frac{\partial \rho}{\partial t}+v\frac{\partial \rho}{\partial x}-\rho\frac{\partial p(\rho)}{\partial x}\right)=0 \Leftrightarrow$$

$$\frac{\partial p(\rho)}{\partial t}+v\frac{\partial p(\rho)}{\partial x}-\rho\frac{\partial p(\rho)}{\partial \rho}\frac{\partial p(\rho)}{\partial x}=0 \Leftrightarrow \quad (19)$$

$$\rho^2\frac{\partial p(\rho)}{\partial \rho}\frac{\partial p(\rho)}{\partial x}=\rho\left(\frac{\partial p(\rho)}{\partial t}+v\frac{\partial p(\rho)}{\partial x}\right)$$

By substituting $\rho^2\frac{\partial p(\rho)}{\partial \rho}\frac{\partial p(\rho)}{\partial x}=\rho\left(\frac{\partial p(\rho)}{\partial t}+v\frac{\partial p(\rho)}{\partial x}\right)$ into the last equation of (18), we get:

$$\frac{\partial(\rho v)}{\partial t}+\frac{\partial(\rho v^2)}{\partial x}+\rho\left(\frac{\partial p(\rho)}{\partial t}+v\frac{\partial p(\rho)}{\partial x}\right)=0 \Leftrightarrow$$

$$\frac{\partial(\rho v)}{\partial t}+\frac{\partial(\rho v^2)}{\partial x}+\rho\left(\frac{\partial p(\rho)}{\partial t}+v\frac{\partial p(\rho)}{\partial x}\right)+p(\rho)\left(\frac{\partial \rho}{\partial t}+\frac{\partial \rho v}{\partial x}\right)=0 \Leftrightarrow$$

$$\frac{\partial(\rho v)}{\partial t}+\frac{\partial(\rho p(\rho))}{\partial t}+\frac{\partial(\rho v^2)}{\partial x}+\frac{\partial(\rho v p(\rho))}{\partial x}=0 \Leftrightarrow \quad (20)$$

$$\frac{\partial(\rho(v+p(\rho)))}{\partial t}+\frac{\partial(\rho v(v+p(\rho)))}{\partial x}=0$$

So, we arrived at the second equation of (19), QED. We can follow the same step backward to get the system (18) from the system (19) using the same expression for the relative velocity of disturbances propagation in the traffic flow: $c(\rho)=-\rho\frac{\partial p(\rho)}{\partial \rho}$.

*2.2 Computational method*

The second order model system (4) has a hyperbolic type, and a variety of finite-difference methods for solving such systems exist. By introducing vectors $W=[\rho,v]^T$ and $f=\left[f_0,\frac{\partial v}{\partial \rho}f_0\right]^T$, the system (4) can be written in the vector form:

$$\frac{\partial W}{\partial t}+A\frac{\partial W}{\partial x}=f \quad (21)$$

with the Jacobi matrix:

$$A = \begin{bmatrix} v & \rho \\ 0 & v + \rho \dfrac{\partial v}{\partial \rho} \end{bmatrix} = \begin{bmatrix} v & \rho \\ 0 & v + c(\rho) \end{bmatrix} \quad (22)$$

For system (4) the whole family of difference schemes could be expressed as:

$$W_m^{n+1} = W_m^n - \frac{\Delta t}{\Delta x}\left(G_{m+1/2}^{n+1/2} - G_{m-1/2}^{n+1/2}\right) + f_m^n \quad (23)$$

In (23), $m = 1,\ldots, M$ is the grid node index along $x$-axis, $n = 1,\ldots, N$ is the grid node index along $t$-axis, $\Delta t$ and $\Delta x$ are numerical integration steps.

The choice of the interpolation method for $G_{m\pm1/2}^{n+1/2}$ in (23) is crucial for obtaining the scheme with the desired properties. When choosing a difference scheme, we must consider that the solution for the model equations at every road segment is defined by the change of computed values in its boundary points. In this case, we can choose the conservative monotonic characteristic method of first-order approximation (Magomedov and Kholodov,1969) with the following expressions for $G_{m\pm1/2}^{n+1/2}$:

$$G_{m\pm1/2}^{n+1/2} = \frac{1}{2} A_{m\pm1/2}^n \left(W_m^n + W_{m\pm1}^n\right) \pm \frac{1}{2}\left(\Omega^{-1}|\Lambda|\Omega\right)_{m\pm1/2}^n \left(W_m^n - W_{m\pm1}^n\right) \quad (24)$$

or these could be more complicated expressions, which allow construction of higher-order accuracy difference schemes on the given stencil. In (24), $\Lambda$ is a diagonal matrix of eigenvalues of Jacobi matrix $A$ and, respectively, $\Omega$ is the matrix of $A$ left eigenvectors and $\Omega^{-1}$ is inverse to $\Omega$. The variable values in intermediate nodes $m\pm1/2$ in (24) can be computed using simple linear interpolation without loss of accuracy.

$$A_{m\pm1/2}^n = \begin{bmatrix} \dfrac{v_m^n + v_{m\pm1}^n}{2} & \dfrac{\rho_m^n + \rho_{m\pm1}^n}{2} \\ 0 & \dfrac{v_m^n + v_{m\pm1}^n}{2} + c\left(\dfrac{\rho_m^n + \rho_{m\pm1}^n}{2}\right) \end{bmatrix}$$

$$\Lambda_{m\pm1/2}^n = \begin{bmatrix} \lambda_{1,m\pm1/2}^n & 0 \\ 0 & \lambda_{2,m\pm1/2}^n \end{bmatrix} = \begin{bmatrix} \dfrac{v_m^n + v_{m\pm1}^n}{2} & 0 \\ 0 & \dfrac{v_m^n + v_{m\pm1}^n}{2} + c\left(\dfrac{\rho_m^n + \rho_{m\pm1}^n}{2}\right) \end{bmatrix}$$

$$\Omega_{m\pm1/2}^n = \begin{bmatrix} \omega_{1,m\pm1/2}^n \\ \omega_{2,m\pm1/2}^n \end{bmatrix} = \begin{bmatrix} \dfrac{2}{\rho_m^n + \rho_{m\pm1}^n} c\left(\dfrac{\rho_m^n + \rho_{m\pm1}^n}{2}\right) & -1 \\ 0 & 1 \end{bmatrix} \quad (25)$$

$$\left(\Omega^{-1}\right)_{m\pm1/2}^n = \begin{bmatrix} \dfrac{\rho_m^n + \rho_{m\pm1}^n}{2c\left(\dfrac{\rho_m^n + \rho_{m\pm1}^n}{2}\right)} & \dfrac{\rho_m^n + \rho_{m\pm1}^n}{2c\left(\dfrac{\rho_m^n + \rho_{m\pm1}^n}{2}\right)} \\ 0 & 1 \end{bmatrix}$$

The characteristic form of compatibility equations, equivalent to the systems (21), along the characteristics $\lambda_{1,2} = dx/dt$, can be expressed as:

$$\omega_{1,2} \cdot \left( \frac{\partial W}{\partial t} + \lambda_{1,2} \frac{\partial W}{\partial x} \right) = \omega_{1,2} \cdot f \qquad (26)$$

Each of two equations (26) is the ordinary differential equation along the corresponding characteristic $\lambda_{1,2} = dx/dt$. Sometimes we need to numerically integrate the compatibility conditions (26) at boundary points along the characteristics entering the integration area. For example, at the beginning of the freeway segment at boundary point $(t^{n+1}, x_1)$ with negative eigenvalue $(\lambda_2)_{3/2}^n = \frac{v_1^n + v_2^n}{2} + c\left(\frac{\rho_1^n + \rho_2^n}{2}\right) < 0$ we use:

$$\begin{cases} (\omega_2)_{3/2}^n \cdot \left( \frac{W_1^{n+1} - W_1^n}{\Delta t} + (\lambda_2)_{3/2}^n \frac{W_2^n - W_1^n}{\Delta x} \right) = (\omega_2)_{3/2}^n \cdot f_1^n \\ v_1^{n+1} = v(t^{n+1}, x_1) \end{cases} \qquad (27)$$

together with boundary condition for traffic velocity $v(t, x_1)$.

At the end of the freeway segment, we do the same for the opposite sign of $(\lambda_2)_{M-1/2}^n = \frac{v_M^n + v_{M-1}^n}{2} + c\left(\frac{\rho_M^n + \rho_{M-1}^n}{2}\right) > 0$:

$$(\omega_{1,2})_{M-1/2}^n \cdot \left( \frac{W_M^{n+1} - W_M^n}{\Delta t} + (\lambda_{1,2})_{M-1/2}^n \frac{W_M^n - W_{M-1}^n}{\Delta x} \right) = (\omega_{1,2})_{M-1/2}^n \cdot f_1^n \qquad (28)$$

for both compatibility equations because $(\lambda_1)_{m\pm 1/2}^n = \frac{v_m^n + v_{m\pm 1}^n}{2} > 0$ always positive.

One can define the relative velocity of disturbances propagation in the traffic flow $c(\rho)$ in (23) and (27),(28) by using equilibrium velocity $V(\rho)$ like it was done in work (Zhang,2002). The function for equilibrium velocity $V(\rho)$ can be set up empirically using traffic detectors data for each segment of the road network.

We propose the new approach without using equilibrium velocity $V(\rho)$ or any form of a fundamental diagram $Q(\rho)$. We just approximate the value of the relative velocity of disturbances propagation $c(\rho)$ by using current observational data for traffic density and velocity:

$$c(\rho)_{m\pm 1/2}^n = \left(\rho \frac{\partial v}{\partial \rho}\right)_{m\pm 1/2}^n = \frac{\rho_m^n + \rho_{m\pm 1}^n}{2} \left( \frac{v_m^n - v_{m\pm 1}^n}{\rho_m^n - \rho_{m\pm 1}^n} \right) \qquad (29)$$

This approach lets our model (4) instantly adjust to the changing situation on the road during simulations.

## 3. Numerical results

To verify the proposed approach, we carried out numerical experiments using traffic detector data for the segment of I-580 freeway in California, USA, obtained from PeMS (California Department of Transportation, 2012). We used two loop detectors, denoted #1 and #2, separated by ~ 1 km long segment of the 4-lane freeway with no entrances or exits, as shown in Fig. 1. The equilibrium velocity $V(\rho)$ and relative velocity of disturbances propagation $c(\rho) = \rho \frac{\partial V(\rho)}{\partial \rho}$ as the functions of density for both detectors together with the observed data for the one-year period are shown in Fig. 2a for the first lane and on Fig. 2b for all four lanes together.

Data (traffic flow and velocity, with 30-second temporal resolution) from detector #1 were used as the left boundary condition, and data from downstream detector #2 were used for verification of modeling results. We simulated a 24-hour interval for a single weekday for all lanes of I-580. The results are shown in Fig. 3 for the first lane and in Fig. 4 for all four lanes together in different subplots: top left — traffic intensities; top right — intensity relative error in logarithmic scale; bottom left — traffic velocities; bottom right — velocity relative error in logarithmic scale. The simulation results obtained with a different expression for the relative velocity of disturbances propagation are shown in different colors: green color — proposed a macroscopic second-order model (4) with $c(\rho) = \rho \frac{\partial v}{\partial \rho}$ from (29); blue color — the same model with $c(\rho) = \rho \frac{\partial V(\rho)}{\partial \rho}$ proposed in (Zhang, 2002). A grey dashed line shows the reference results (data of detector #2).

The simulation results exhibit good agreement but not full match between the results, obtained using the same second-order macroscopic model (4), and the results obtained with a different form of the relative velocity of disturbances propagation in traffic flow $c(\rho)$.

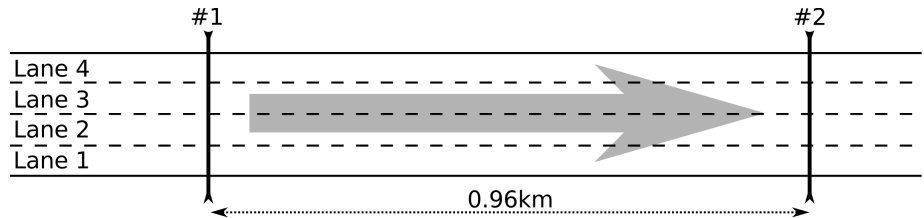

Figure 1: Traffic detectors #1 and #2 chosen on the segment of the I-580 freeway in California, USA. Data from detector #1 were used as a left boundary condition for modeling, and data from downstream detector #2 were used for verification of modeling results.

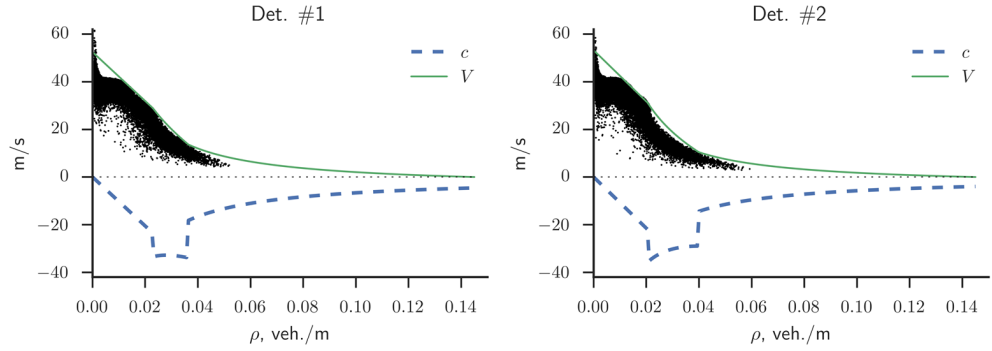

Figure 2a: Equilibrium speed (solid green line) and relative velocity of disturbances (dashed blue line) as the functions of density for the first lane of I-580. Black dots indicate observed values of speed. Left: detector #1, right: detector #2.

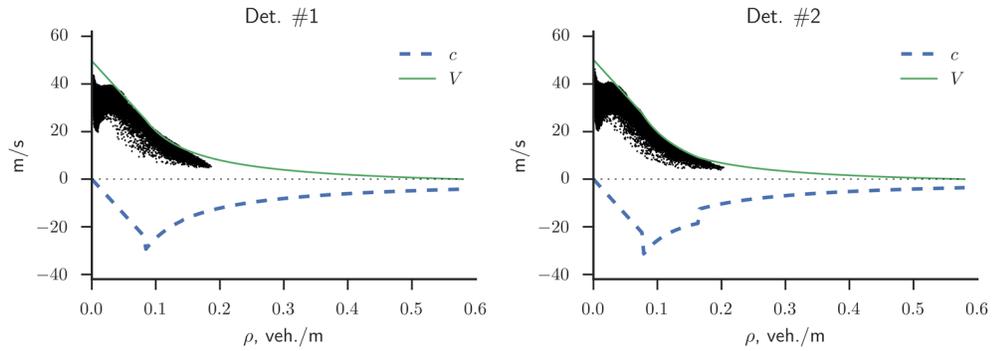

Figure 2b: Equilibrium speed (solid green line) and relative velocity of disturbances propagation (dashed blue line) as the functions of density for all four lanes of I-580. Black dots indicate observed values of speed. Left: detector #1, right: detector #2.

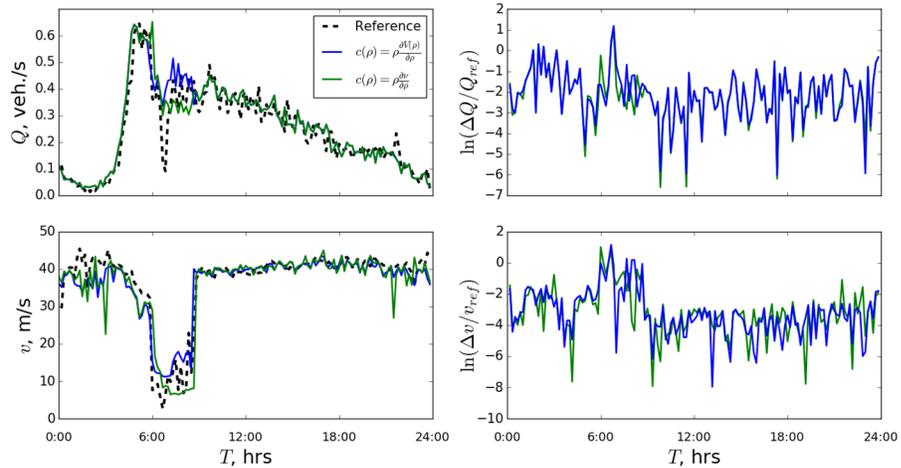

Figure 3: The comparison of calculated intensities (top) and velocities (bottom) from the first lane of I-580 and observed values from detector #2 (dashed grey lines). On the right — relative errors of intensity (top) and velocity (bottom) in logarithmic scale. The simulation results obtained with a different expression for the relative velocity of disturbances propagation $c(\rho)$ are shown in different colors: green color — model (4) with $c(\rho)$ from (29); blue color — the same model with $c(\rho)$ proposed in (Zhang, 2002).

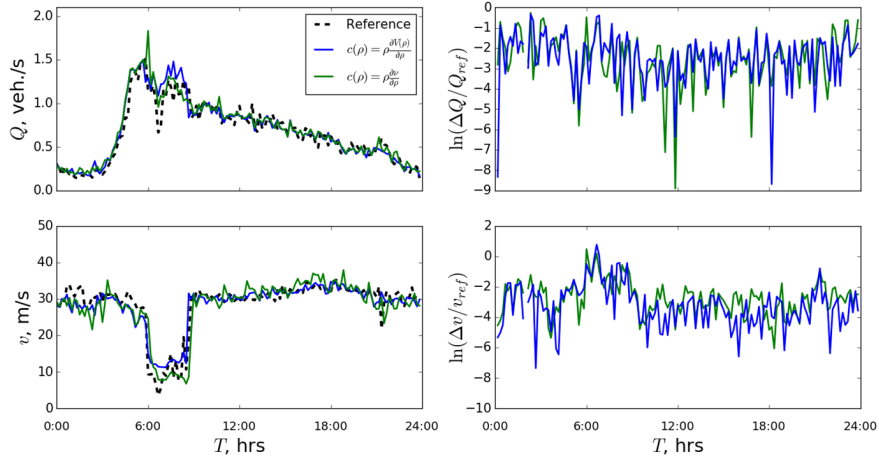

Figure 4: The comparison of calculated intensities (top) and velocities (bottom) aggregated over four lanes of I-580, and observed values from detector #2 (dashed grey lines). On the right — relative errors of intensity (top) and velocity (bottom) in logarithmic scale. The simulation results obtained with a different expression for the relative velocity of disturbances propagation $c(\rho)$ are shown in different colors: green color — model (4) with $c(\rho)$ from (29); blue color — the same model with $c(\rho)$ proposed in (Zhang, 2002).

## 4. Conclusion

In this paper we explored the unification of the second-order hydrodynamic macroscopic traffic models in various formulations (Payne, 1971; Daganzo, 1995; Papageorgiou, 1998; Aw and Rascle, 2000; Zhang, 2002, 2003; Siebel and Mauser, 2006a,b). Existing second-order macroscopic models describe traffic by a non-linear system of the hyperbolic equations (for density and speed), which differ in the way they account for dependency between traffic flow (or velocity) and density. We have shown that all these second-order macroscopic models can be generalized using the expression for the relative velocity of disturbances propagation in the traffic flow.

To verify proposed methodology, we conducted numerical experiments by simulating the segment of I-580 freeway in California, USA, using data from PeMS (California Department of Transportation, 2012). The simulations show that the results obtained for the same macroscopic model using a different form of the relative velocity of disturbances propagation slightly different. This suggests that the properties of every phenomenological model are defined by the form of its dependency form the density for the relative velocity of disturbances propagation in traffic flow.

Also, it is essential that now we know the answer to the question why the transition from macroscopic first-order models to second-order is useful. The answer is simple. The adding of the second equation for velocity into the system allows removing the fundamental diagram from simulations. We just approximate the value of the relative velocity of disturbances propagation using current observational data for traffic density and velocity in (29) and forget about the diagram.

**Supplementary Figures**

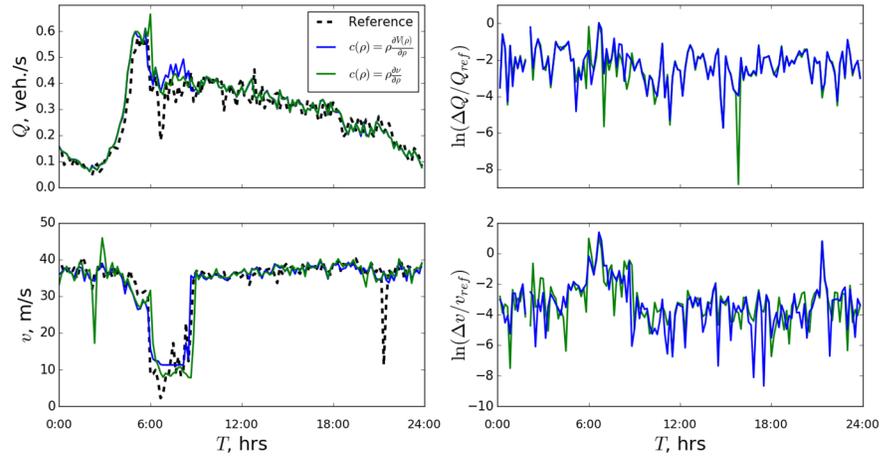

Figure S1: The comparison of calculated intensities (top) and velocities (bottom) for the second lane of I-580, and observed values from detector #2 (dashed grey lines). On the right — relative errors of intensity (top) and velocity (bottom) in logarithmic scale. The simulation results obtained with a different expression for the relative velocity of disturbances propagation $c(\rho)$ are shown in different colors: green color — model(4) with $c(\rho)$ from (29); blue color — the same model with $c(\rho)$ proposed in (Zhang,2002).

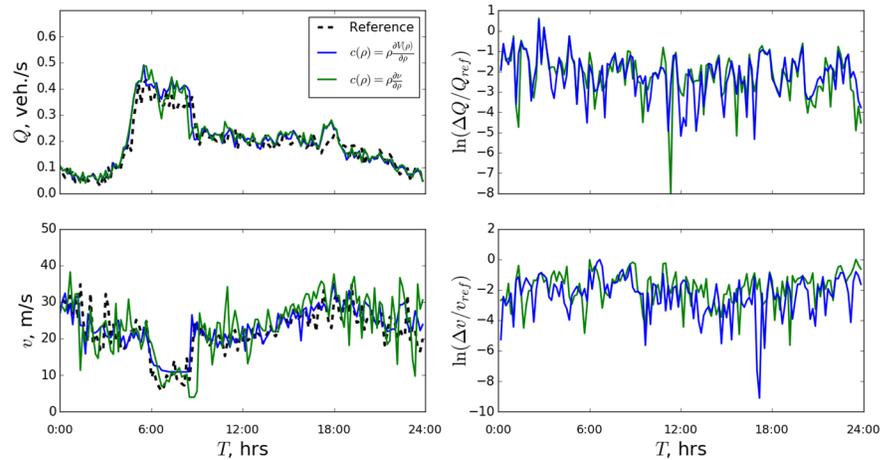

Figure S2: The comparison of calculated intensities (top) and velocities (bottom) for the third lane of I-580, and observed values from detector #2 (dashed grey lines). On the right — relative errors of intensity (top) and velocity (bottom) in logarithmic scale. The simulation results obtained with a different expression for the relative velocity of disturbances propagation $c(\rho)$ are shown in different colors: green color — model(4) with $c(\rho)$ from (29); blue color — the same model with $c(\rho)$ proposed in (Zhang,2002).

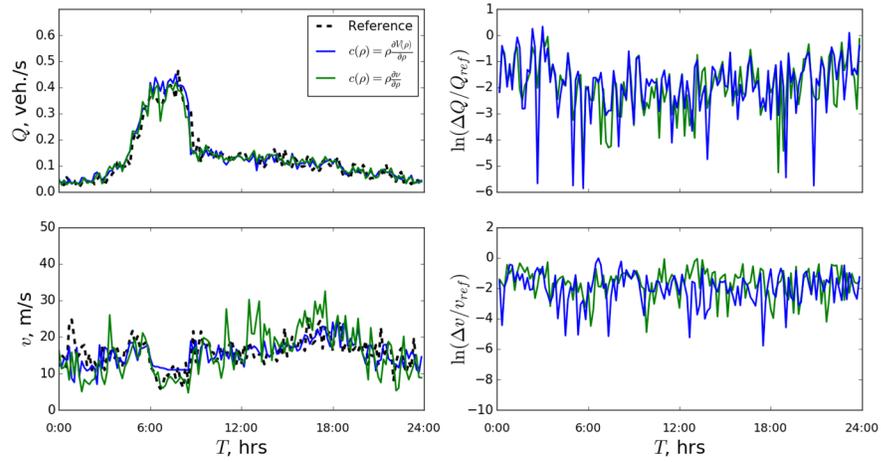

Figure S3: The comparison of calculated intensities (top) and velocities (bottom) for the fourth lane of I-580, and observed values from detector #2 (dashed grey lines). On the right — relative errors of intensity (top) and velocity (bottom) in logarithmic scale. The simulation results obtained with a different expression for the relative velocity of disturbances propagation $c(\rho)$ are shown in different colors: green color — model (4) with $c(\rho)$ from (29); blue color — the same model with $c(\rho)$ proposed in (Zhang, 2002).